\documentclass{article}
\usepackage[T1]{fontenc}
\usepackage{iclr2027_conference,times}
\usepackage{amsmath,amsfonts,bm}

\def\eqref#1{equation~\ref{#1}}
\def\1{\bm{1}}

\DeclareMathAlphabet{\mathsfit}{\encodingdefault}{\sfdefault}{m}{sl}
\SetMathAlphabet{\mathsfit}{bold}{\encodingdefault}{\sfdefault}{bx}{n}

\usepackage{amsmath}
\usepackage{amssymb}
\usepackage{booktabs}
\usepackage{multirow}
\usepackage{graphicx}
\usepackage{subcaption} % 如果子图才需要，单图不用

\usepackage{float} % 用来支持[H]选项，强制图片就在当前位置
\usepackage{algorithm}
\usepackage{algpseudocode}

\usepackage{hyperref}
\usepackage{url}
\usepackage{caption}
\usepackage{xcolor}
\usepackage{colortbl} % 用来实现绿色行高亮
\definecolor{proposalred}{RGB}{178,34,34}
\usepackage{makecell}
\usepackage{fontawesome5}
\usepackage{wrapfig}
\definecolor{arrowup}{HTML}{B56A6A}   % muted red
\definecolor{arrowdown}{HTML}{5F858D} % muted blue-teal

\title{SEPM: Self-Evolving Procedural Memory Grounded in Task Graphs for Multi-Agent Collaboration%
  \ificlrfinal
  \thanks{This work was conducted during an internship at China Telecom and was supported in part by the China Telecom Key Research Project (No.~65996).}
  \fi
}

\author{%
\makebox[\textwidth][c]{%
\parbox{\textwidth}{%
\centering
Yi Liu$^{1}$ \quad Junwei Zhao$^{2}$ \quad Li Sun$^{1}$ \quad Jiansong Zhang$^{2}$ \quad Sen Su$^{1}$\\[0.5em]
\normalfont
$^{1}$Beijing University of Posts and Telecommunications \quad $^{2}$China Telecom\\[0.5em]
\faEnvelope\ \texttt{\mdseries louie@bupt.edu.cn}
}%
}%
}

\newcommand{\method}{\textsc{SEPM}}

\iclrfinalcopy
\begin{document}

\maketitle
\begin{abstract}
Long-horizon multi-agent collaboration requires memory that preserves both successful actions and the coordination structure underlying them. Existing memory systems support information sharing or procedural reuse, but often leave the evolution of shared procedures disconnected from the task dependencies and execution states of collaborating agents. This coupling also limits the direct transfer of procedural memory designed for individual agents to team settings. We propose Self-Evolving Procedural Memory (SEPM), a framework that uses task dependency graphs as a common representation for online coordination and procedural memory evolution. During execution, a lightweight shared blackboard records structured evidence alongside task dependencies, supporting the detection of goal, plan, and world-state divergence and evidence-based realignment. After execution, verified actions and dependencies are distilled into reusable procedures, while recurring failures inform warnings and constraints. The shared procedure pool evolves through creation, refinement, specialization, and retirement, guided by success and cost utility and constrained by evidence and safety requirements. Together, these mechanisms preserve the collaborative structure and applicability conditions needed to reuse experience across tasks. Evaluations on ALFWorld, WebArena, and OfficeBench under AutoGen and DyLAN show that SEPM achieves the highest task scores among the compared methods in all six benchmark settings. Under AutoGen, it improves task scores over the no-added-memory baseline by 29.85, 25.86, and 13.00 percentage points, respectively. The codebase is available at \url{https://github.com/cookieApril/SEPM}.
\end{abstract}

\section{Introduction}
\label{sec:intro}

External memory enables LLM agents to reuse experience through reflections, skills, and procedural workflows \citep{shinn2023reflexion,wang2024voyager,wang2025awm,fang2025memp}. A natural route from individual agents to teams is to retain these memory units and add mechanisms for sharing, retrieval, and role-specific allocation. This adaptation-first route provides a bridge from single-agent memory to multi-agent reuse. However, making a procedure available to multiple agents does not establish the conditions under which their separate actions compose into a valid collaborative execution.

In a team, a remembered step may depend on an artifact produced by another agent, a verified state change, or a completed handoff. These conditions must remain consistent as agents act and revise their plans. When memory records what to do but leaves these dependencies to be reconstructed during execution, agents may duplicate work, proceed before prerequisites are satisfied, or reuse procedures whose assumptions no longer hold. Such failures motivate treating procedural memory and coordination as a coupled design problem: the dependencies that make experience reusable are also the dependencies that must be maintained during collaboration.

Recent multi-agent memory systems organize collaborative experience through role-specific procedures, hierarchical graphs, and team-level memory \citep{han2026legomem,zhang2025gmemory,wu2026scaling}. We pursue a coordination-native route in which collaborative dependencies form the interface between persistent memory and ongoing execution. The design question is how one representation can preserve the conditions of successful collaboration across tasks while making those conditions available for checking and correcting the current task.

We propose Self-Evolving Procedural Memory (\method{}), which unifies multi-agent memory and coordination through task dependency graphs. Nodes represent executable steps, and directed edges encode prerequisites between them. Retrieved procedures are instantiated into task graphs linked to agent assignments and verified execution state on a shared blackboard. During execution, these dependencies support plan comparison, prerequisite checks, and the propagation of corrections to affected agents. The resulting evidence also determines what the team remembers: successful repairs revise the active procedure, improvements simplify or strengthen its executable structure, and conditional lessons are attached as warnings. Previously unseen workflows create new entries. The same dependency semantics therefore connect the procedure that guides collaboration to the evidence that revises it.

\noindent
\begin{minipage}[t]{0.46\textwidth}
We evaluate \method{} across four benchmarks to examine how coupling procedural memory with shared collaboration state improves multi-agent performance. On the coding, research, and database scenarios of MultiAgentBench~\citep{zhu2025multiagentbench}, \method{} achieves the highest Task Score (TS) and Communication Score (CS) among the compared methods in each scenario, as shown in Figure~\ref{fig:env_perf}. We evaluate \method{} on ALFWorld, WebArena, and OfficeBench under AutoGen and DyLAN. It achieves the highest task score among the compared methods in all six benchmark settings. Under AutoGen, replacing explicit dependency graphs with textual Procedural Memory while retaining procedural information and the same retrieval rules and context budgets reduces task scores by 17.16, 8.13,
\end{minipage}
\hfill
\begin{minipage}[t]{0.5\textwidth}
% 关键：消除figure[H]上下默认间距 \intextsep
\setlength{\intextsep}{0pt}
\setlength{\abovecaptionskip}{1pt}
\setlength{\belowcaptionskip}{0pt}
\begin{figure}[H]
\centering
% 向上挪图片，压缩顶部空白；负数向上移动
\vspace{-4pt}
\includegraphics[width=\linewidth]{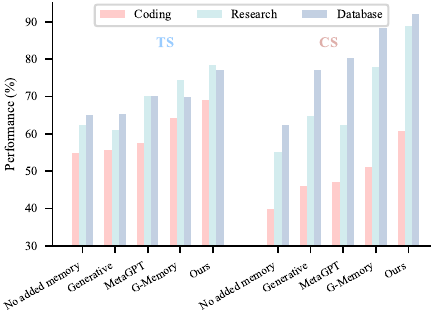} 
\caption{Task Score (TS) and Communication Score (CS) across the coding, research, and database environments of MultiAgentBench under AutoGen.}
\label{fig:env_perf}
\end{figure}
\end{minipage}
\vspace{-0.3pt}
and 12.34 percentage points, respectively. This comparison supports the contribution of explicit dependency graphs within the proposed architecture. On ALFWorld, \method{} also achieves higher task performance than Generative Agents and G-Memory with lower total evaluation token use.
 
Our contributions can be summarized as follows:

\textbf{(i)} First, we formulate a {coordination-native approach to procedural memory}, using task dependencies to connect reusable experience with ongoing collaboration.

\textbf{(ii)} Second, we develop a {shared graph representation} that supports online alignment and the extraction and revision of collaborative procedures.

\textbf{(iii)} Third, we introduce a {divergence-aware alignment mechanism} that uses this shared state to detect goal, plan, and world-state inconsistencies.

\section{Related Work}

\paragraph{From single-agent procedures to multi-agent reuse.}
Single-agent procedural memory stores reusable skills and workflows derived from execution experience\citep{wang2024voyager,wang2025awm,fang2025memp, NEURIPS2024_8ac50fd0,NEURIPS2025_9549f7d0}. One route to multi-agent reuse transfers these units into a team and adds role-specific access or allocation\citep{han2026legomem}. LEGOMem adapts the single-agent Synapse and AWM baselines to task agents, while its own method constructs modular full-task and subtask memories \citep{han2026legomem}. This transfer route makes existing experience available to multiple agents, but the allocation of a procedure alone does not specify when a cross-agent prerequisite has been established. \method{} makes these prerequisites explicit and uses them in both collaborative execution and subsequent memory revision.

\paragraph{Multi-agent memory architectures and graph semantics.}
G-Memory organizes experience through interaction, query, and insight graphs \citep{zhang2025gmemory}. LEGOMem distributes procedural knowledge across orchestration and
task execution \citep{han2026legomem}. LLMA-Mem combines episodic, procedural, and transactive memory for lifelong team learning \citep{wu2026scaling}. These systems establish multi-agent memory as an architectural design problem \citep{wang2025mirix,hao2026decentmem,yu2026architecture}. \method{} studies a specific coupling between memory and coordination: task-graph edges express prerequisites between executable steps, and the same dependency semantics are used to instantiate procedures, compare plans, propagate corrections, and retain supported execution structure. This distinguishes the proposed use of graphs from organizing experience primarily for retrieval\citep{xu2025amem,feng2026expgraph}.

\paragraph{Shared workspaces and coordination.}
Blackboard architectures provide shared state for exchanging intermediate results and coordinating work \citep{HAYESROTH1985251,nii1995blackboard,wu2024autogen, han2025blackboard,chen2024agentverse,hong2024metagpt, salemi2025blackboard}. \method{} uses this architecture to associate public execution records with task nodes, agent assignments, and
supporting evidence. The records describe execution outcomes, while the dependency graph identifies which later actions rely on them. Their association allows an alignment correction to inform both the current plan and the procedural guidance available to subsequent tasks.

\section{Problem Formulation}
\label{sec:formulation}

We consider a team $\mathcal{A}=\{a_1,\ldots,a_n\}$ coordinated by a manager agent over a sequence of tasks $\tau_1,\tau_2,\ldots$. Each agent maintains private memory, while the team shares a task-specific blackboard and a procedural memory pool $\mathcal{P}_j$. We study how collaborative execution can improve this pool while preserving the task dependencies and execution constraints required for subsequent reuse.

\paragraph{Shared execution state.}
At step $t$ of task $\tau$, the shared workspace and its task graph are
\begin{equation}
W_\tau^t
=
\bigl(g_\tau^\star,\mathcal{G}_\tau^t,B_\tau^t,\Sigma_\tau^t\bigr),
\qquad
\mathcal{G}_\tau^t
=
\bigl(V_\tau^t,E_\tau^t,X_\tau^t,Z_\tau^t\bigr).
\label{eq:workspace}
\end{equation}
$g_\tau^\star$ specifies the objective and constraints; $B_\tau^t$ stores public execution records; and $\Sigma_\tau^t$ contains verified task-relevant state at the current stage of task execution. The directed acyclic graph $\mathcal{G}_\tau^t$ represents logical subtasks through nodes $V_\tau^t$, prerequisites through edges $E_\tau^t$, and action descriptions through $X_\tau^t$. The associated execution metadata $Z_\tau^t$ records agent assignments and step status. Retry and replanning semantics are given in Appendix~\ref{app:dag_retries}.

\paragraph{Procedural memory.}
A versioned procedure is $S_i^{(v)}=(M_i^{(v)},\Pi_i^{(v)},G_i^{(v)},\mathcal{W}_i^{(v)})$. The metadata $M_i^{(v)}$ contains the title, description, applicability conditions, tags, context conditions, exclusions and version identifier. The step set $\Pi_i^{(v)}$ stores the instruction for each executable step, while the directed acyclic graph $G_i^{(v)}=(V_i^{(v)},E_i^{(v)})$ encodes their prerequisite relations. $\mathcal{W}_i^{(v)}$ contains conditional warnings together with their triggering conditions and supporting evidence. 

\paragraph{Sequential memory revision.}
For each reusable procedure, the pool retains an active revision $S_i^{(v)}$. The active revision serves as the reference baseline when a new task either reveals limitations of an existing procedure or presents an opportunity to improve it. A candidate revision is assessed against this reference using the recorded outcomes, costs, and task constraints relevant to the update. We write the resulting pool transition as
\begin{equation}
\mathcal{P}_{j+1}
=\mathcal{U}(\mathcal{P}_j,H_{\tau_j}),
\label{eq:memory_revision}
\end{equation}
where $H_{\tau_j}$ contains the task's execution records, dependency structure, verified state, and outcome. The update operator $\mathcal{U}$ creates a procedure for a new workflow, revises an existing procedure when execution supports a repair or improvement, or attaches conditional lessons to its warnings. Structural changes preserve the prerequisites and mandatory checks needed for valid collaboration. This formulation makes memory evolution an outcome of collaborative execution: the task graph both guides the current team and identifies which changes can be retained for future reuse.

\begin{figure}[t]
  \centering
  \vspace{-35pt}
  \captionsetup{skip=1pt}
  \includegraphics[width=0.98\linewidth]{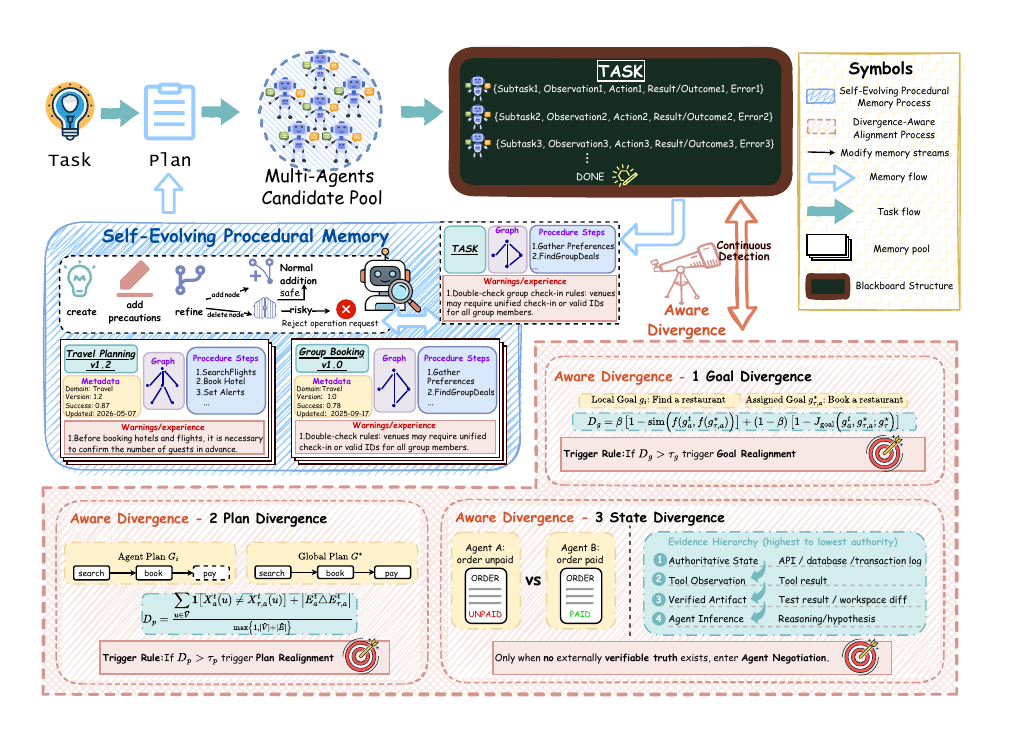}
  \caption{Overview of \method{}. Retrieved procedures instantiate collaborative task graphs. Execution records linked to task nodes support goal, plan, and state alignment. Verified outcomes and prerequisite relations support procedure creation, structural revision, and updates to conditional warnings.}
  \label{fig:framework}
  \vspace{-10pt}
\end{figure}

\section{\method}
\label{sec:method}

\method{} couples procedural reuse and collaborative execution through a shared blackboard, divergence-aware alignment, and graph-grounded memory revision. As illustrated in Figure~\ref{fig:framework}, retrieved procedural memories guide the decomposition and planning of target tasks. During execution, unmet prerequisites or plan deviations are identified and remediated, and these verified adjustments are then integrated into the shared procedure pool for reuse in future tasks.

\subsection{Shared Blackboard for Procedural Execution}
\label{sec:blackboard}

\paragraph{Retrieval and task instantiation.}
For a task query $q$ with verified context facts $F_q$, retrieval considers active procedures whose context conditions are contained in $F_q$ and
whose exclusions do not intersect $F_q$. Let $\mathcal{C}(q,F_q)$ denote this
candidate set. We rank candidates by
\begin{equation}
\begin{aligned}
s_i(q)
&=w_b(q)\,\widetilde{\mathrm{BM25}}(q,S_i)
+w_v(q)\,\operatorname{sim}\bigl(f(q),f(S_i)\bigr),\\
\mathcal{R}(q)
&=\operatorname{TopK}_{S_i\in\mathcal{C}(q,F_q)}s_i(q),
\qquad K=5,\quad w_b(q)+w_v(q)=1.
\end{aligned}
\label{eq:retrieval}
\end{equation}
Here, $f$ denotes the text representation function. The retrieval configuration and adaptive weighting rule are provided in Appendix~\ref{app:retrieval_defaults}. The manager binds the retrieved steps to task entities and agent responsibilities while retaining applicable prerequisites and warnings.

\paragraph{Structured execution.}
During execution, agent $a$ contributes
\begin{equation}
b_{a,t}
=
(q_{a,t},o_{a,t},u_{a,t},r_{a,t},\epsilon_{a,t}),
\qquad
B_\tau^{t+1}=B_\tau^t\oplus b_{a,t},
\label{eq:execution_record}
\end{equation}
corresponding to the assigned subtask, observation, executed action, result, and error. Each entry is linked to its graph node. Private reasoning remains local. The manager updates node status and incorporates verified observations into $\Sigma_\tau^t$. These records preserve both successful actions and failures needed for subsequent procedure extraction.

\subsection{Divergence-Aware Multi-Agent Alignment}
\label{sec:alignment}

For agent $a$, let $(g_a^t,\mathcal{G}_a^t,\Sigma_a^t)$ denote its local objective, plan, and state estimate. Its reference objective $g_{\tau,a}^\star$ and subgraph $\mathcal{G}_{\tau,a}^t$ are determined by the current assignment, including prerequisites supplied by other agents. Alignment evaluates consistency with the agent's role.

\begin{figure}[t]
  \centering
  \vspace{-40pt}
  \captionsetup{skip=1pt}
  \includegraphics[width=0.96\linewidth]{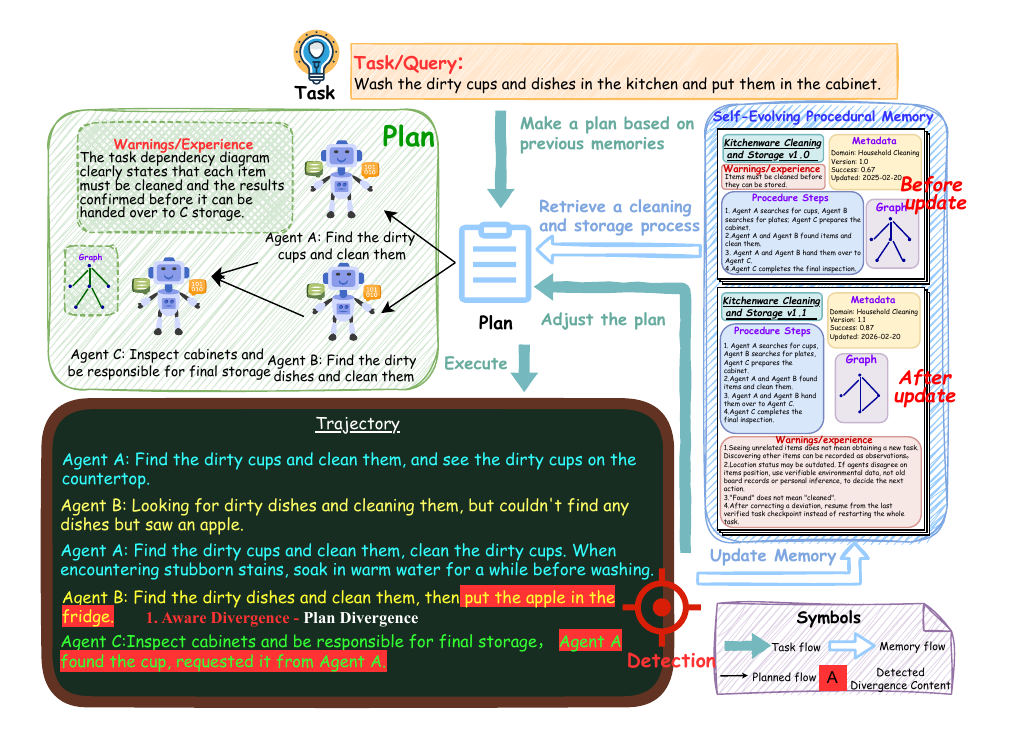}
  \caption{
  Illustrative cleaning-and-storage case. Agent B introduces an unrelated action, while Agent C requests a cup before cleaning has been verified. Realignment restores the assigned plan and the cleaning-before-handoff prerequisite. The revised procedure retains this dependency and adds guidance on state verification and resumption from verified checkpoints.
  }
  \label{fig:case_study}
  \vspace{-10pt}
\end{figure}

\paragraph{Goal divergence.}
We combine a discrepancy between goal-text representations with a task-conditioned goal-consistency score:
\begin{equation}
D_g
=
\beta\left[
1-\operatorname{sim}\bigl(f(g_a^t),f(g_{\tau,a}^\star)\bigr)
\right]
+
(1-\beta)\left[
1-J_{\mathrm{goal}}(g_a^t,g_{\tau,a}^\star;g_\tau^\star)
\right].
\label{eq:goal_divergence}
\end{equation}
$\beta\in[0,1]$ balances the two terms. The first term measures dissimilarity between the textual
representations of the current and assigned goals. The second term, $J_{\mathrm{goal}}\in[0,1]$, assesses whether the intended operation is consistent with the agent's assignment and the global task constraints. This distinction matters when similar goal descriptions imply different operations, such as finding and booking a restaurant. The representation and scoring details are provided in Appendix~\ref{app:goal_divergence}.

\paragraph{Plan divergence.}
Plan divergence measures differences between an agent's local plan and its assigned plan in both actions and dependencies. After matching corresponding nodes, let $\bar V=V_a^t\cup V_{\tau,a}^t$ and $\bar E=E_a^t\cup E_{\tau,a}^t$. We define
\begin{equation}
D_p
=
\frac{
\sum_{u\in\bar V}
\mathbf{1}\!\left[X_a^t(u)\neq X_{\tau,a}^t(u)\right]
+
\left|E_a^t\triangle E_{\tau,a}^t\right|
}{
\max\left\{1,|\bar V|+|\bar E|\right\}
}.
\label{eq:plan_divergence}
\end{equation}
The node-action term $\sum_{u\in\bar V}\mathbf{1}[X_a^t(u)\neq X_{\tau,a}^t(u)]$ counts action mismatches between corresponding nodes, while the edge term $|E_a^t\triangle E_{\tau,a}^t|$ counts dependency relations that occur in only one plan. Their sum is normalized by the combined plan size, so larger $D_p$ indicates greater plan divergence. Node matching and missing-node handling are detailed in Appendix~\ref{app:alignment_details}.

\paragraph{State divergence and realignment.}
For each task-relevant fact $k$, let $\mathcal{E}_k^+$ denote the qualified
evidence reported by the agents. We define
\begin{equation}
\begin{aligned}
C_s
&=
\left\{
k:
\exists e,e'\in\mathcal{E}_k^+,\;
\operatorname{val}(e)\neq\operatorname{val}(e')
\right\},\\
z_a^t
&=
\mathbf{1}
[D_g>\tau_g\;\lor\;D_p>\tau_p\;\lor\;C_s\neq\varnothing].
\end{aligned}
\label{eq:realignment_trigger}
\end{equation}
 $C_s$ contains task-relevant facts for which the agents provide conflicting qualified evidence. When $z_a^t=1$, the manager resolves the detected goal, plan, or state conflict and propagates the correction to affected agents. Evidence qualification and conflict resolution are detailed in Appendix~\ref{app:state_resolution}.

Figure~\ref{fig:case_study} illustrates how these signals lead to different corrections. Agent A is assigned to clean cups, Agent B to clean dishes, and Agent C to inspect the cabinet and complete storage. Agent B's action of placing an apple in the refrigerator introduces an action outside its assigned plan. The correction redirects B to the dish-cleaning subtask. Agent C's request for a cup raises a different issue: locating the cup establishes its presence, but does not establish that cleaning is complete. The handoff therefore requires evidence of the cleaning outcome before storage proceeds. The dependency links A's verified completion to C's next action, while cabinet inspection can proceed independently. Thus, realignment concerns both the actions assigned to each agent and the evidence required for a cross-agent handoff.

\subsection{Self-Evolving Procedural Memory}
\label{sec:evolution}

\paragraph{Extraction from verified collaboration.}
At task completion, let $H_\tau=(B_\tau^{T_\tau},\mathcal{G}_\tau^{T_\tau}, \Sigma_\tau^{T_\tau},y_\tau)$ contain the terminal records, graph, verified state, and outcome. The reusable execution structure is
\begin{equation}
\begin{aligned}
V_\tau^+
&=
\{u\in V_\tau^{T_\tau}:H_\tau\models u\},\\
E_\tau^+
&=
\{(u,w)\in E_\tau^{T_\tau}:
u,w\in V_\tau^+,\ H_\tau\models(u,w)\}.
\end{aligned}
\label{eq:supported_graph}
\end{equation}

$H_\tau\models u$ indicates that the recorded action and its outcome are verified. For a dependency $(u,w)$, the required support concerns whether the outcome of $u$ establishes a prerequisite of $w$; temporal precedence alone does not establish this relation. Extraction $\widehat S_\tau=\Phi(H_\tau,V_\tau^+,E_\tau^+)$ converts supported actions into reusable steps, retains their prerequisite relations, and associates recurring failures with conditional warnings or preconditions. Appendix~\ref{app:extraction_verification} describes procedure extraction and revision.

\subsection{Revision Validation and Promotion}
\label{app:revision_rules}

\paragraph{Candidate validation and promotion.}
Execution histories provide candidate procedures and revisions. Candidate validation screens the proposed change, while promotion determines whether it becomes an active procedure version. Let $\mathcal{T}_{S'}$ denote the recorded reproduction trials for candidate $S'$. Promotion requires
\begin{equation}
\begin{aligned}
\operatorname{Accept}(S';\mathcal{T}_{S'})
={}&G_{\mathrm{safety}}
\land G_{\mathrm{state}}
\land G_{\mathrm{causal}}\\
&\land G_{\mathrm{reproduction}}
\land [\overline{u}(S')\geq0].
\end{aligned}
\label{eq:revision_acceptance}
\end{equation}
The gates check safety, verified state, causal attribution, and reproduction requirements. The mean net benefit $\overline{u}(S')$ is computed over all recorded trials, including failed trials. A candidate without recorded trials cannot be promoted. Appendix~\ref{app:extraction_verification} specifies the candidate checks, trial utility, and promotion summary score.

\begin{figure}[t]
\centering
\vspace{-35pt}
\begin{subfigure}[t]{0.485\linewidth}
    \centering
    \includegraphics[width=\linewidth]{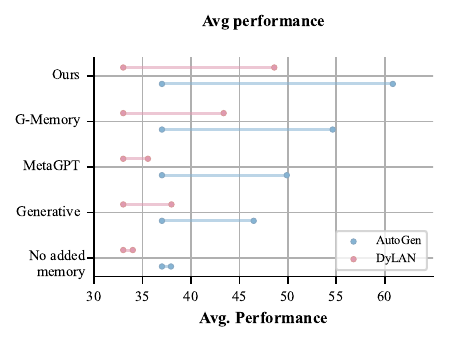}
    \caption{Average task score by host.}
    \label{fig:avg_gain}
\end{subfigure}\hfill
\begin{subfigure}[t]{0.485\linewidth}
    \centering
    \includegraphics[width=\linewidth]{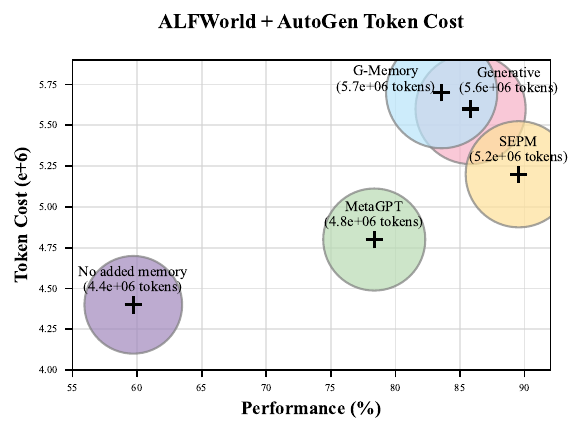}
    \caption{ALFWorld performance versus token cost.}
    \label{fig:token_cost_alfworld_autogen}
\end{subfigure}
\caption{
Task performance and token cost.
(a) Mean task scores across ALFWorld, WebArena, and OfficeBench;
each segment connects the no-added-memory baseline to a memory strategy
within the same host.
(b) ALFWorld task scores versus total input and output tokens over the complete evaluation under AutoGen, measured in millions.
}
\label{fig:performance_cost_pair}
\vspace{-10pt}
\end{figure}

\section{Experiments}
\label{sec:experiments}

We evaluate whether graph-grounded procedural memory improves collaborative task performance across two hosts. We then examine sensitivity to the joint choice of execution and memory models, the contributions of the blackboard, procedural memory, and alignment, and the effect of replacing explicit dependency graphs with textual procedures. Token consumption is measured alongside task performance to characterize the cost of the complete evaluation.

\subsection{Experimental Setup}
\label{sec:experimental_setup}

ALFWorld evaluates embodied procedural interaction \citep{shridhar2021alfworld}; WebArena evaluates stateful web tasks \citep{zhou2023webarena}; and OfficeBench evaluates workflows spanning office applications \citep{wang2024officebench}. We report benchmark-specific task scores using the evaluation criteria of each environment, with higher scores indicating better task performance. 

\begin{wraptable}{r}{0.58\linewidth}
\vspace{-1.1\baselineskip}
\centering
\renewcommand{\arraystretch}{0.9}
\footnotesize
\setlength{\tabcolsep}{2.5pt}

\newcommand{\upd}[1]{\textcolor{arrowup}{\scriptsize$\uparrow$#1}}
\newcommand{\dnd}[1]{\textcolor{arrowdown}{\scriptsize$\downarrow$#1}}

\caption{Task performance across host multi-agent systems. Boldface marks the highest score within each benchmark setting. Arrows indicate percentage-point changes relative to the same host without added memory.}
\label{tab:main}

\begin{tabular}{@{}llrrr@{}}
\toprule
MAS & Memory & ALFWorld & WebArena & OfficeBench \\
\midrule

\multirow{5}{*}{\rotatebox[origin=c]{90}{AutoGen}}
& No added memory & 59.70\upd{0.00} & 12.44\upd{0.00} & 41.67\upd{0.00} \\
& Generative      & 85.82\upd{26.12} & 15.27\upd{2.83} & 38.33\dnd{3.34} \\
& MetaGPT         & 78.36\upd{18.66} & 25.99\upd{13.55} & 45.33\upd{3.66} \\
& G-Memory        & 83.58\upd{23.88} & 30.30\upd{17.86} & 50.00\upd{8.33} \\
& \method{}       & \textbf{89.55}\upd{29.85} & \textbf{38.30}\upd{25.86} & \textbf{54.67}\upd{13.00} \\

\midrule

\multirow{5}{*}{\rotatebox[origin=c]{90}{DyLAN}}
& No added memory & 53.73\upd{0.00} & 14.90\upd{0.00} & 33.33\upd{0.00} \\
& Generative      & 65.67\upd{11.94} & 15.27\upd{0.37} & 33.00\dnd{0.33} \\
& MetaGPT       & 49.25\dnd{4.48} & 22.41\upd{7.51} & 35.00\upd{1.67} \\
& G-Memory        & 61.94\upd{8.21} & 25.49\upd{10.59} & 42.67\upd{9.34} \\
& \method{}       & \textbf{68.66}\upd{14.93} & \textbf{29.80}\upd{14.90} & \textbf{47.33}\upd{14.00} \\
\bottomrule
\end{tabular}
\vspace{-1.0\baselineskip}
\end{wraptable}

The host comparison includes no added memory, Generative Agents \citep{park2023generative}, MetaGPT \citep{hong2024metagpt}, G-Memory \citep{zhang2025gmemory}, and \method{} under AutoGen \citep{wu2024autogen} and DyLAN \citep{liu2024a}. The main planning, task-execution, and procedural memory models, as well as the goal-consistency judge, use \texttt{gpt-5-mini} in the main comparison. The sensitivity analysis keeps AutoGen, the main planning agent, and the goal-consistency judge fixed while jointly varying the task-execution and procedural memory model over twelve alternative endpoints. The component analysis progressively introduces the shared blackboard, procedural memory, and divergence-aware alignment. The final comparison reports total input and output token consumption recorded during the corresponding ALFWorld evaluations under AutoGen. Appendix~\ref{app:protocol} documents the evaluation scope and reporting details.

We additionally evaluate the five memory configurations on the coding, research, and database environments of MultiAgentBench. Appendix~\ref{app:benchmark_protocols} specifies the evaluation configuration and scoring protocol used to assess performance across these environments.

\subsection{Performance across Host Systems}
\label{sec:host_comparison}

\method{} achieves the highest observed task score in all six benchmark settings (Table~\ref{tab:main}). Relative to the strongest alternative in each setting, it improves ALFWorld, WebArena, and OfficeBench by 3.73, 8.00, and 4.67 percentage points under AutoGen, and by 2.99, 4.31, and 4.66 points under DyLAN. The consistent ranking across hosts supports the usefulness of the proposed memory coordination design under different orchestration mechanisms.

The margins nevertheless vary across tasks and hosts. The largest advantage appears on WebArena under AutoGen and on OfficeBench under DyLAN, indicating that the benefit depends on how the host executes a particular class of workflows. The baseline results show a related pattern: Generative Agents performs best among the alternatives on ALFWorld, whereas G-Memory leads on WebArena and OfficeBench. Thus, the comparison favors different memory designs across environments, while \method{} remains effective across all three.

Figure~\ref{fig:avg_gain} summarizes the overall improvement: mean task score rises from 37.94\% to 60.84\% under AutoGen and from 33.99\% to 48.60\% under DyLAN. The component and representation analyses below examine how shared execution state, procedural evolution, and explicit dependencies contribute to these gains.

\subsection{Sensitivity to Execution and Memory Models}
\label{sec:model_sensitivity}

We evaluate sensitivity to the joint choice of task-execution and procedural memory model.
AutoGen, the main planning agent, and the goal-consistency judge remain fixed, while each listed endpoint is used for both task execution and procedural memory management. Table~\ref{tab:procedure_model} includes the main configuration from Table~\ref{tab:main} as a reference, together with twelve alternative endpoints.

\begin{wraptable}{r}{0.45\linewidth}
\vspace{-1.29\baselineskip}
\centering
\caption{Sensitivity to execution and memory models. The main planning agent and goal-consistency judge are fixed. AW, WA, and OB denote ALFWorld, WebArena, and OfficeBench, respectively.}
\label{tab:procedure_model}
% \vspace{-0.2cm}
\footnotesize
\renewcommand{\arraystretch}{1.05}
\setlength{\tabcolsep}{2.9pt}
\begin{tabular}{@{}lccc@{}}
\toprule
Model & AW & WA & OB \\
\midrule
qwen3.5-0.8b & 58.21 & 14.53 & 45.67 \\
qwen3.5-2b & 57.46 & 14.78 & 45.67 \\
qwen3.5-9b & 67.16 & 16.01 & 49.67 \\
qwen3.5-27b & 81.34 & 30.42 & 55.67 \\
gemma-4-12b-it & 72.39 & 22.54 & 53.00 \\
gemma-4-31b-it & 77.61 & 28.69 & 54.33 \\
\midrule
deepseek-v4-flash-0731 & \textbf{93.28} & 38.05 & 64.33 \\
qwen3.8-max & 89.55 & 39.41 & 63.67 \\
glm-5.2 & 85.82 & 37.93 & 60.67 \\
kimi-k3 & 90.30 & 42.36 & 65.33 \\
\midrule
gpt-5-mini (main) & 89.55 & 38.30 & 54.67 \\
gpt-5.6-terra & 91.79 & \textbf{43.23} & \textbf{67.67} \\
claude-opus-5 & 92.54 & 42.49 & 65.33 \\
\bottomrule
\end{tabular}
\vspace{-1.5\baselineskip}
\end{wraptable}

Performance varies substantially with the joint execution-and-memory model, with score ranges of 35.82, 28.70, and 22.00 percentage points on ALFWorld, WebArena, and OfficeBench. Within the Qwen3.5 configurations, increasing model size from 2B to 27B improves all benchmarks, whereas the 0.8B and 2B configurations perform similarly. These results suggest that the framework depends on the model's ability to execute procedural guidance and maintain useful memories, with limited gains between the two smallest configurations.

The strongest endpoint differs by benchmark. \texttt{deepseek-v4-flash-0731} achieves the highest ALFWorld score, while \texttt{gpt-5.6-terra} leads on WebArena and OfficeBench. Relative to the main configuration, the latter improves these benchmarks by 2.24, 4.93, and 13.00 points, respectively. The uneven gains indicate that model choice has different consequences across task environments. Since execution and memory management change together, this experiment measures their joint effect.

\subsection{Component Contributions}
\label{sec:components}

Figure~\ref{fig:graph_ablation_components} evaluates procedural memory and divergence-aware alignment on a shared blackboard. The blackboard provides the task dependency graph and structured execution records on which both mechanisms operate. We therefore retain this structural foundation when evaluating the contribution of either mechanism. 

The shared blackboard improves task scores by 9.70, 15.27, and 3.33 percentage points on ALFWorld, WebArena, and OfficeBench, respectively. Its largest gain occurs on WebArena, suggesting that maintaining a common execution state is particularly useful when successive actions interact with a changing web environment.

Building self-evolving memory on this foundation yields an additional 17.17 points on ALFWorld, 6.28 on WebArena, and 0.33 on OfficeBench. The large ALFWorld gain is consistent with the value of retaining reusable action sequences and their prerequisites. On OfficeBench, however, procedural reuse alone provides little improvement over the blackboard, suggesting that access to a workflow is insufficient when its execution depends on coordinated state changes across applications.

Alignment changes this pattern. Adding it to Blackboard + memory improves OfficeBench by 9.34 points, compared with 2.98 on ALFWorld and 4.31 on WebArena. Conversely, the OfficeBench gain from memory increases from 0.33 points without alignment to 3.67 points with alignment. This pattern suggests that maintaining execution consistency helps the team realize the benefit of reusable office procedures. The full configuration performs best on all three benchmarks, with task-dependent contributions from memory evolution and online alignment.

\subsection{Explicit Graphs versus Textual Procedures}
\label{sec:graph_control}

\begin{figure}[t]
  \centering
  \vspace{-25pt}
  % 设置为略大于左右内容中较高者的高度
  \def\panelheight{4.0cm}

  % (a) 左侧
  \begin{subfigure}[t]{0.49\linewidth}
    \centering
    \begin{minipage}[c][\panelheight][c]{\linewidth}
      \centering
      \includegraphics[width=\linewidth]{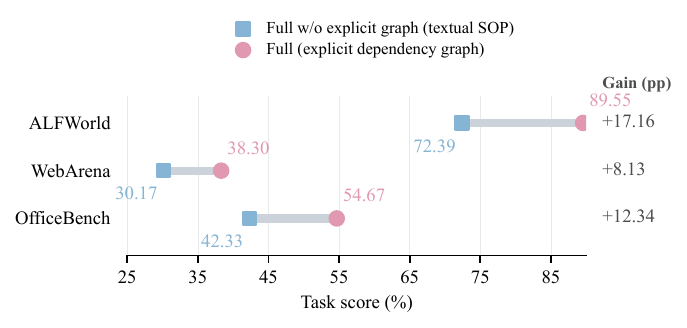}
    \end{minipage}
    \vspace{-8pt}
    \caption{Comparison with textual procedural memory.}
    \label{fig:graph_ablation_comparison}
  \end{subfigure}
  \hfill
  % (b) 右侧
  \begin{subfigure}[t]{0.49\linewidth}
    \centering
    \begin{minipage}[c][\panelheight][c]{\linewidth}
      \centering
      \footnotesize
      \renewcommand{\arraystretch}{1.18}
      \setlength{\tabcolsep}{2pt}
      \begin{tabular}{@{}lccc@{}}
        \toprule
        Condition & AW & WA & OB \\
        \midrule
        No added components    & 59.70 & 12.44 & 41.67 \\
        Blackboard only        & 69.40 & 27.71 & 45.00 \\
        Blackboard + memory    & 86.57 & 33.99 & 45.33 \\
        Blackboard + alignment & 82.09 & 34.48 & 51.00 \\
        Full \method{}         & \textbf{89.55} &
                                 \textbf{38.30} &
                                 \textbf{54.67} \\
        \bottomrule
      \end{tabular}
    \end{minipage}
    \vspace{-8pt}
    \caption{Component ablation.}
    \label{fig:graph_ablation_components}
  \end{subfigure}

  \caption{
    Ablation results under AutoGen.
    \textbf{(a)} Full \method{} versus the textual procedural memory
    control.
    \textbf{(b)} Component configurations. AW, WA, and OB denote
    ALFWorld, WebArena, and OfficeBench. Memory denotes the complete
    cross-task procedure extraction, revision, and reuse mechanism built
    on the shared blackboard.
  }
  \label{fig:graph_ablation}
  \vspace{-10pt}
\end{figure}

We introduce Full w/o explicit graph, graph objects and adjacency relations are replaced by textual procedures that state each step's prerequisites. This intervention changes the dependency representation and its associated operations without introducing additional task information or model capabilities.Replacing explicit graphs with textual prerequisites reduces task scores by 17.16, 8.13, and 12.34 percentage points on ALFWorld, WebArena, and OfficeBench, respectively (Figure~\ref{fig:graph_ablation_comparison}). Because the control retains procedural content and prerequisite descriptions, the performance gap suggests that preserving this information in memory is insufficient to obtain the full benefit of the proposed design.

Explicit edges make the same dependency structure available to prerequisite checking, plan comparison, and correction propagation. A change to one step can therefore be related directly to the downstream steps that depend on it. The textual control requires these relations to be recovered from descriptions during the corresponding operations. The results support the combined use of explicit dependencies and graph-based execution operations for turning procedural knowledge into coordinated behavior.

\subsection{Cost analysis}
\method{} achieves 89.55\% success using 5.2 million tokens over the 134 ALFWorld tasks (Figure~\ref{fig:token_cost_alfworld_autogen}). Relative to no added memory, token consumption increases by approximately 18.2\%, while task success improves by 29.85 percentage points. The additional computation therefore accompanies a larger fraction of completed tasks.

Compared with Generative Agents and G-Memory, \method{} improves success by 3.73 and 5.97 points while reducing total token consumption by approximately 7.1\% and 8.8\%, respectively. It thus achieves a better trade-off between performance and token consumption than both memory baselines in this evaluation. MetaGPT consumes fewer tokens but reaches lower success, representing a different trade-off. These results show that the additional coordination and memory-update mechanisms can coexist with lower overall token consumption than the two competing memory systems.

\section{Conclusion and Limitations}
\label{sec:conclusion}

\paragraph{Conclusion.}
We introduced \method{}, which connects procedural memory with multi-agent coordination through shared task dependencies. The blackboard grounds execution in a common graph, alignment maintains consistency during collaboration, and procedural memory evolves through execution-supported revisions across tasks. Procedural memory provides a large gain on ALFWorld, while OfficeBench benefits more from its combination with alignment. Under AutoGen, \method{} achieves task scores of 89.55\%, 38.30\%, and 54.67\% on ALFWorld, WebArena, and OfficeBench, respectively, outperforming the information-matched textual control on all three benchmarks. On ALFWorld, it also uses fewer tokens than Generative Agents and G-Memory.

% \paragraph{Limitations.}
% Performance depends on the execution and memory-management model, and the strongest model differs across benchmarks. Model selection therefore remains an important consideration when applying the framework to new workflows. Our evaluation lacks a dedicated benchmark for jointly assessing multi-agent memory and collaboration. Our next step is to develop a benchmark specifically for multi-agent memory and collaboration, with controlled variation in task recurrence and coordination requirements to better characterize their influence on collaboration and memory benefits.
\paragraph{Limitations.}
The current SEPM implementation represents collaborative procedures as directed acyclic graphs of executable steps and explicit prerequisite relations. This formulation targets workflow-oriented collaboration in which task dependencies and relevant execution states can be identified during interaction. More open-ended settings may contain iterative or less explicit coordination structures. Incorporating richer temporal and dependency representations is a natural extension of the framework and could broaden its applicability to these forms of collaboration.

\subsection*{AI Use Statement}
Generative AI assisted with prose revision, result-table formatting, visualization code, and consistency checks against the supplied manuscript, implementation, figures, and experiment records.

\subsection*{Ethics Statement}
This work studies procedural memory and coordination for language-model agents in established benchmark environments. The experiments do not involve human participants or the collection of new personal data, and the evaluated agents are not deployed in user-facing or consequential real-world settings. The framework may inherit biases, unsafe behavior, and factual errors from its underlying language models; its memory and alignment mechanisms should not be interpreted as guarantees of safety or correctness.

\subsection*{Reproducibility Statement}
We provide the algorithmic specification of \method{}, including its execution flow, retrieval and alignment rules, procedure extraction and verification, retry behavior, and version-update mechanism, in Appendices~\ref{app:execution_flow}--\ref{app:versioning}. Section~\ref{sec:experimental_setup} and Appendix~\ref{app:protocol} document the evaluated host systems, model configurations, comparison conditions, memory initialization and persistence, benchmark environments, task counts, execution budgets, random-seed construction, score aggregation, and token accounting. Appendix~\ref{app:additional_results} reports supplementary results and additional analyses. We will publicly release the implementation, benchmark adapters, experiment configurations, and evaluation scripts to support reproduction of the reported results.

\bibliography{references}
\bibliographystyle{iclr2027_conference}

\clearpage

\appendix

\section{Algorithm Overview}
\label{app:execution_flow}

Algorithm~\ref{alg:sepm_revision} summarizes how \method{} connects collaborative execution to procedural memory evolution. Retrieved procedures initialize a task dependency graph; structured execution records support alignment during the task and memory updates after it.

\begin{algorithm}[H]
\caption{SEPM: collaborative execution and procedural memory evolution}
\label{alg:sepm_revision}
\begin{algorithmic}[1]
\Require Task $\tau$, procedure pool $\mathcal{P}$, execution budget $\mathcal{B}_{\mathrm{exec}}$
\Ensure Task outcome $y$ and updated procedure pool $\mathcal{P}$
\State $\mathcal{R}\gets\operatorname{RetrieveCompatible}(\tau,\mathcal{P})$
\State $(\mathcal{G},B,\Sigma)\gets\operatorname{InitializeTask}(\tau,\mathcal{R})$
\While{task unfinished and budget remains}
    \State Execute a ready step, or obtain missing evidence and replan
    \State Append $(\mathrm{task},\mathrm{observation},\mathrm{action},\mathrm{result},\mathrm{error})$ to $B$
    \State Update step status and resolve verified state $\Sigma$ from the recorded evidence
    \If{goal, plan, or state divergence is detected}
        \State Realign affected agents and update $\mathcal{G}$ and its prerequisites
    \EndIf
\EndWhile
\State $y\gets\operatorname{Outcome}(\tau,B,\Sigma)$; $H\gets(B,\mathcal{G},\Sigma,y)$
\For{each prechecked candidate $S'$ with reference $S$ proposed from $H$}
    \State $\mathcal{T}_{S'}\gets\operatorname{RecordedTrials}(S')$
    \If{$\operatorname{Accept}(S';\mathcal{T}_{S'})$}
        \State Create or atomically revise the active procedure in $\mathcal{P}$
    \EndIf
\EndFor
\State Atomically attach supported conditional lessons to the relevant procedures
\State \Return $y,\mathcal{P}$
\end{algorithmic}
\end{algorithm}

\paragraph{Execution and update operations.}
A ready step has verified prerequisites in the current task state. When no step is ready, the host controller obtains the missing observation or revises the plan. Execution, observation, retries, and replanning use the task's existing budget. The outcome $y$ records completion or the terminal failure or unresolved status. Each retrieved procedure retains the revision used as the reference for its update.

$\operatorname{ProposeUpdates}$ extracts supported steps and dependencies from $H$ to form new procedures or revisions. Both new procedures and revisions require trial-based promotion under Equation~\ref{eq:revision_acceptance}. Trial records are supplied through the reproduction-recording interface; candidates without sufficient promotion evidence remain outside the active pool. Updates retain their applicability conditions, and conditional lessons are stored with their supporting evidence. Warning updates can be retained even when the task does not support a structural revision. The commit operation preserves revision history as described in Appendix~\ref{app:versioning}.

\section{Method Details}
\label{app:method_details}

\subsection{Procedure Retrieval}
\label{app:retrieval_defaults}

Retrieval filters the active procedure pool by context conditions and exclusions, then ranks compatible entries using Equation~\ref{eq:retrieval}. The default configuration retrieves $K=5$ procedures. BM25 uses $k_1=1.5$ and $b=0.75$; its scores are normalized by the maximum in the candidate set. The default text representation $f$ uses 256-dimensional BLAKE2b-based signed feature hashing with $\ell_2$ normalization and a bounded cosine-based similarity. Indexed fields include the title, description, task family, applicability, tags, context conditions, warnings, and step instructions.

Adaptive retrieval initializes the BM25 and vector logits to zero and applies a softmax to obtain their weights. The implementation's routing cues modify these logits as follows:

\begin{table}[t]
\centering
\small
\caption{Query-dependent logit adjustments for adaptive procedure retrieval.}
\label{tab:retrieval-routing}
\begin{tabular}{@{}lll@{}}
\toprule
Query cues & Modified logit & Increment \\
\midrule
\texttt{code}, \texttt{debug}, \texttt{implement}, \texttt{coding} & Vector & $1.0$ \\
\texttt{latest}, \texttt{current}, \texttt{real-time} & BM25 & $0.8$ \\
\texttt{routine}, \texttt{repeat}, \texttt{standard} & BM25 & $0.6$ \\
\bottomrule
\end{tabular}
\end{table}

With no routing cue, both weights are $0.50$. Disabling adaptive retrieval also fixes $w_b=w_v=0.50$.

\subsection{Goal and Plan Alignment}
\label{app:alignment_details}

\paragraph{Goal comparison.}
\label{app:goal_divergence}
Goal divergence combines similarity between hashed text representations with task-conditioned goal consistency. The representation term uses the mapping $f$ described in Appendix~\ref{app:retrieval_defaults}. For $J_{\mathrm{goal}}$, \texttt{gpt-5-mini} receives the agent's current goal, its assigned goal, and the global task objective with its constraints. It returns a score in $[0,1]$, where higher values indicate greater consistency of the intended operation with the assignment and global constraints. We set $\beta=0.55$ and trigger goal realignment when $D_g>0.35$.

\paragraph{Plan comparison.}
\label{app:plan_divergence}
Nodes are matched by persistent subtask identifiers assigned during task-graph instantiation. A different action field at a matched identifier counts as one node edit; a missing node receives a null action label. Edge differences are the symmetric difference of the directed edge sets. Ownership and completion status are stored separately and do not enter the score. Equation~\ref{eq:plan_divergence} therefore measures a normalized discrepancy under fixed node correspondence. Plan realignment is triggered when $D_p>0.40$.

\subsection{Evidence-Based State Resolution}
\label{app:state_resolution}

A state key $k$ identifies an entity-attribute pair. Each evidence item $e$ records a value, authority tier $h(e)$, source $s(e)$, observation time $t(e)$, and confidence $c(e)$. Evidence is eligible when $c(e)\geq0.70$:
\begin{equation}
\mathcal{E}_k^+=\{e\in\mathcal{E}_k:c(e)\geq0.70\}.
\label{eq:evidence_resolution}
\end{equation}
If this set is empty, the state remains unresolved and the controller requests an observation. Otherwise, the resolver selects the highest available tier,
\begin{equation}
h_k^\star=\max_{e\in\mathcal{E}_k^+}h(e),
\qquad
\mathcal{E}_k^\star=\{e\in\mathcal{E}_k^+:h(e)=h_k^\star\}.
\end{equation}
The authority order is environment or official state, tool observation, verified artifact, and agent inference. Agreement within the highest qualified tier establishes the reported value. For conflicting observations from the same source and tier, the latest observation is used; a tie remains unresolved. Conflicts across sources require a fresh authoritative or tool observation. Confidence determines eligibility rather than breaking ties between contradictory values.

Agent inference alone cannot establish a verified external state. The controller first seeks external evidence and uses negotiation when no external source is available. A newer report that conflicts with stored state triggers re-observation of the relevant key. State resolution operates on individual keys and same-source observation times; it does not implement global temporal reasoning or a time-to-live policy.

\subsection{Task Graphs, Retries, and Replanning}
\label{app:dag_retries}

The task graph represents the current prerequisite structure. Each logical step has a persistent identifier and an action description. Self-loops, duplicate edges, and cyclic dependencies are rejected. Repeating an action creates another execution record for its logical step rather than a backward edge in the graph.

After a failed check, the controller may retry the step or the manager may produce a revised acyclic plan. Unchanged steps retain their identifiers; new steps receive new identifiers. A failed or modified step requires fresh verification of affected downstream conditions, while unaffected prerequisites retain their supporting evidence. Execution ends when the task completes, the host budget is exhausted, or the controller returns an unresolved state.

\subsection{Procedure Extraction and Revision}
\label{app:extraction_verification}

The terminal history $H_\tau$ associates blackboard records with the task graph, verified state, and outcome. Extraction uses the supported node and edge sets in Equation~\ref{eq:supported_graph}: a retained node has evidence for its action and outcome, and a retained edge has evidence for the corresponding prerequisite relation. The resulting procedure abstracts supported steps and retains their dependencies, applicability conditions, and conditional warnings.

Execution of a previously unseen workflow can produce a candidate procedure, while experience with an existing procedure can produce a candidate revision. Candidates remain separate from the active procedure pool until recorded trials satisfy the promotion requirements. Execution failures and other supported observations may also be retained as conditional lessons. Recording a candidate or an execution event does not itself create an active procedure version. Failures can yield conditional warnings without promoting an unverified structural change. Retries and repairs are part of the current execution.

\paragraph{Candidate prechecks.}
At the proposal and validation stages, a candidate without recorded trials is screened using its supplied benefit and cost:
\begin{equation}
u_{\mathrm{claim}}
=b_{\mathrm{claim}}-\lambda c_{\mathrm{claim}},
\qquad \lambda=1.0.
\end{equation}
This quantity is a candidate-level estimate used for prechecking. It does not replace the reproduction evidence required for promotion.

\paragraph{Trial-based promotion.}
For candidate $S'$, let $\mathcal{T}_{S'}$ contain all recorded reproduction trials. Each trial $t$ supplies a candidate reward, a baseline reward, and a cost:
\begin{equation}
u_t
=r_t^{\mathrm{candidate}}
-r_t^{\mathrm{baseline}}
-\lambda c_t.
\end{equation}
Promotion evaluates the mean net benefit
\begin{equation}
\overline{u}(S')=
\begin{cases}
\displaystyle
\frac{1}{|\mathcal{T}_{S'}|}
\sum_{t\in\mathcal{T}_{S'}}u_t,
& |\mathcal{T}_{S'}|>0,\\[6pt]
-\infty,
& |\mathcal{T}_{S'}|=0.
\end{cases}
\end{equation}
Failed trials are included in this average. The baseline reward is supplied in each trial record for the reference procedure or no-candidate baseline under the same validation setting. It is not inferred from the aggregate benchmark tables.

Promotion requires the safety, state-verification, causal-attribution, reproduction, and nonnegative-net-benefit gates to pass. The default safety-risk threshold is $0.50$, the causal-confidence threshold is $0.65$, and the minimum reproduction count is $1$.

\paragraph{Validation summary score.}
After successful promotion, the created procedure version stores
\begin{equation}
v_{\mathrm{summary}}
=
\min\left\{
1,\;
0.4c_{\mathrm{causal}}
+
0.3\frac{n_q}{\max(m,n_q)}
+
0.3(1-r_{\mathrm{safety}})
\right\},
\end{equation}
where $c_{\mathrm{causal}}$ is the candidate's supplied causal confidence, $n_q$ is the number of qualified trials, $m$ is the minimum reproduction count, and $r_{\mathrm{safety}}$ is the safety-risk score. With the default $m=1$, the reproduction term equals $0.3$ whenever at least one trial qualifies. The summary score is stored as version metadata and is not a separate promotion gate.

\subsection{Atomic Version Updates}
\label{app:versioning}

An accepted structural or warning update creates a new revision $v'$. The update becomes active only if the current head still matches the revision $v$ used to construct the candidate:
\begin{equation}
\bigl(S_i^{(v)},\operatorname{head}_i=v\bigr)
\xrightarrow[\operatorname{head}_i=v]{\text{accepted update}}
\bigl(S_i^{(v')},\operatorname{head}_i=v'\bigr).
\label{eq:version_commit}
\end{equation}
A failed head comparison leaves the active revision unchanged. Superseded revisions remain in history, while retrieval uses the active revision.

\section{Experimental Protocols}
\label{app:protocol}

\subsection{Hosts, Models, and Comparison Conditions}

The main comparison evaluates no added memory, Generative Agents, MetaGPT, G-Memory, and \method{} under AutoGen and DyLAN. Each host retains its native orchestration and within-task context. The main planning, execution, and procedural-memory models, together with the goal-consistency judge, use \texttt{gpt-5-mini}. The model-sensitivity experiment fixes AutoGen, the main planning agent, and the goal-consistency judge while jointly changing the execution and procedural-memory endpoint. The benchmark tasks, execution budgets, and evaluators remain unchanged across these configurations. Table~\ref{tab:procedure_model} contains twelve alternative endpoints and the main configuration.

The component experiment compares the host baseline, blackboard only, blackboard with procedural memory, blackboard with alignment, and the full system. The textual-procedure control retains the blackboard, memory, alignment, retrieval rules, context budget, and update mechanism. It expresses prerequisites in text and uses textual references for plan comparison and correction propagation in place of explicit graph objects and adjacency relations.

\subsection{Memory Initialization and Persistence}
\label{app:memory_protocol}

Each benchmark condition run starts with an empty procedure pool. Tasks within the run use the same SQLite database, allowing stored execution records, candidates, trials, and accepted procedure versions to persist across the task sequence. Separate experimental conditions use separate database paths. A resumed run continues from its existing database and checkpoint rather than initializing a new pool.

The generic runtime uses an output-specific database path, which can be overridden through \texttt{TEAM\_MEMORY\_DB}. At agent or task start, the adapter requests up to $K=5$ compatible procedures. Context-specific variants are eligible only when their context conditions are satisfied by the facts supplied to the adapter.

Tasks follow the sequence selected by the benchmark adapter. For ALFWorld, each task resets the environment, and its case seed is
\[
s_i=s_{\mathrm{base}}\times1000003+i,
\]
where $i$ is the official task index. WebArena and OfficeBench use their respective runtime initialization and reset procedures.

\subsection{Benchmark Environments and Evaluation}
\label{app:benchmark_protocols}

\paragraph{ALFWorld.}
We evaluate 134 tasks in the text-based environment with a maximum of 30 actions per task. Evaluation uses the task feedback returned by the ALFWorld environment. The environment resets between tasks, while the memory mechanism retains experience across the evaluated sequence.

\paragraph{WebArena.}
We evaluate 812 tasks through BrowserGym's action interface using accessibility-tree observations. Tasks are executed in the sequence selected by the WebArena adapter. Each method updates memory from completed interactions using its own update mechanism. Task scores are obtained from WebArena's functional evaluator.

\paragraph{OfficeBench.}
OfficeBench~\citep{wang2024officebench} evaluates workflows involving documents, spreadsheets, email, and calendar operations. We evaluate all 300 tasks, comprising 93 single-application, 95 two-application, and 112 three-application tasks. The environment runs in Docker and provides tool APIs organized into Word, Excel, Calendar, Email, System, and OCR-PDF capabilities~\citep{han2026legomem}. Each task specifies an initial collection of files and application data on which the agents operate. After execution, task-specific evaluators inspect the resulting files, application state, or submitted answers against the expected outcome. These checks include exact matching and fuzzy keyword matching where required by the task. We aggregate the task-level evaluation scores over all 300 tasks.
\paragraph{MultiAgentBench.}
We evaluate 100 tasks in each of the coding, research, and database environments~\citep{zhu2025multiagentbench,wu2026scaling}. All five conditions use AutoGen, with \texttt{gpt5.5-mini} for execution and \texttt{claude-opus-5} for evaluation. Task definitions, agent profiles, environment rules, and communication graphs are retained. Each task permits at most three iterations, and memory persists across tasks within a condition's sequence. The generation temperature is $0.7$ with a maximum sequence length of $1024$; coding uses $0.1$ and $4096$, respectively.

Task Score (TS) measures output quality or correctness; Communication Score (CS) measures communication and planning. Research judgments of innovation, safety, and feasibility are normalized into TS by the benchmark evaluator. Scores retain the benchmark's reporting scale. The communication graph specifies which agents communicate, while the \method{} task graph specifies prerequisites between executable steps.

\subsection{Evaluation Units and Aggregation}
\label{app:aggregation}

Each task condition pair is evaluated once with base seed $0$. The evaluation includes 134 ALFWorld tasks, 812 WebArena tasks, and 300 OfficeBench tasks. We retain the task-level scores returned by the corresponding benchmark evaluation pipeline and aggregate them over the evaluated task set.

Figure~\ref{fig:avg_gain} reports an unweighted arithmetic mean of the three displayed benchmark scores for each method configuration. Differences between displayed scores are reported in points on the same reporting scale. MultiAgentBench TS and CS are reported separately and are excluded from this mean.

\subsection{Token Accounting}
\label{app:token_accounting}

Token consumption is the sum of input and output tokens over the complete 134-task ALFWorld evaluation under AutoGen. Each total is paired with the task score from the same run in Table~\ref{tab:main}. The reported totals are $4.4$, $5.6$, $4.8$, $5.7$, and $5.2$ million tokens for no added memory, Generative Agents, MetaGPT, G-Memory, and \method{}, respectively.

\section{Supplementary Results and Analysis}
\label{app:additional_results}

\subsection{Task Quality and Communication on MultiAgentBench}

\begin{table}[t]
\centering
\caption{Task and communication scores on MultiAgentBench. All methods use AutoGen, \texttt{gpt5.5-mini} for execution,
and \texttt{claude-opus-5} for evaluation. TS and CS retain the benchmark's reporting scale.}
\label{tab:multiagentbench_scores}
\small
\setlength{\tabcolsep}{5pt}
\begin{tabular}{@{}lrrrrrr@{}}
\toprule
& \multicolumn{2}{c}{Coding} & \multicolumn{2}{c}{Research}
& \multicolumn{2}{c}{Database} \\
\cmidrule(lr){2-3}\cmidrule(lr){4-5}\cmidrule(lr){6-7}
Method & TS & CS & TS & CS & TS & CS \\
\midrule
No added memory & 54.90 & 39.81 & 62.40 & 55.16 & 64.93 & 62.43 \\
Generative & 55.74 & 46.09 & 61.16 & 64.64 & 65.21 & 77.16 \\
MetaGPT & 57.62 & 47.23 & 70.11 & 62.27 & 70.14 & 80.29 \\
G-Memory & 64.20 & 51.17 & 74.29 & 77.77 & 69.73 & 88.20 \\
\method{} & \textbf{69.09} & \textbf{60.64} & \textbf{78.37} & \textbf{88.91} & \textbf{77.03} & \textbf{91.94} \\
\bottomrule
\end{tabular}
\end{table}

Table~\ref{tab:multiagentbench_scores} reports the values underlying Figure~\ref{fig:env_perf}. Relative to G-Memory, \method{} increases TS by $4.89$, $4.08$, and $7.30$ points in coding, research, and database, respectively, and CS by $9.47$, $11.14$, and $3.74$ points. The gains occur in both output and communication scores, which is consistent with the framework's joint treatment of task execution and coordination. Their magnitudes differ across environments: research has the largest CS gain over G-Memory, whereas database has the largest TS gain. These aggregate scores describe complementary outcomes rather than establishing that one causes the other.

\subsection{Conditional Contributions of Memory and Alignment}
\label{app:model_intervention}

The component experiment forms a two-factor comparison conditional on the shared blackboard. Let $s_B$, $s_{BM}$, $s_{BA}$, and $s_{BMA}$ denote scores with the blackboard, memory, and alignment components indicated by the subscripts. The descriptive interaction contrast is
\begin{equation}
I=s_{BMA}-s_{BM}-s_{BA}+s_B.
\end{equation}

\begin{table}[htbp]
\centering
\small
\caption{Conditional gains from the component experiment, in percentage points. Differences are computed from the displayed task scores.}
\label{tab:conditional_gains}
\begin{tabular}{@{}lrrr@{}}
\toprule
Comparison & ALFWorld & WebArena & OfficeBench \\
\midrule
Add memory to blackboard & $17.17$ & $6.28$ & $0.33$ \\
Add memory to blackboard + alignment & $7.46$ & $3.82$ & $3.67$ \\
Add alignment to blackboard & $12.69$ & $6.77$ & $6.00$ \\
Add alignment to blackboard + memory & $2.98$ & $4.31$ & $9.34$ \\
Interaction $I$ & $-9.71$ & $-2.46$ & $3.34$ \\
\bottomrule
\end{tabular}
\end{table}

On ALFWorld and WebArena, each extension has a smaller marginal gain when the other is present, indicating overlapping benefits on the observed score scale. OfficeBench shows the opposite pattern: memory adds $0.33$ points to the blackboard alone but $3.67$ points when alignment is available. Alignment similarly adds more when memory is present. This pattern suggests that the value of reusable office procedures depends in part on maintaining the shared state needed to execute them. The contrast is descriptive and is computed from one run per condition.

\end{document}